\documentclass[11pt,preprint]{aastex}
\usepackage{emulateapj5}
\usepackage{apjfonts}

\submitted{Accepted by the Astrophysical Journal 2004 February 20}
\shortauthors{Boboltz \& Claussen}
\shorttitle{Ground-State SiO Masers Toward Evolved Stars}

\begin{document}

\title{Ground-State ${\rm SiO}$ Maser Emission Toward Evolved Stars} 
\author{D. A. Boboltz}
\affil{U.S. Naval Observatory, \\ 
3450 Massachusetts Ave., NW, Washington, DC 20392-5420 \\
dboboltz@usno.navy.mil}
\and
\author{M. J. Claussen}
\affil{National Radio Astronomy Observatory, \\
P.O. Box O, Socorro, NM 87801 \\
mclausse@nrao.edu}

\begin{abstract}  

We have made the first unambiguous detection of vibrational ground-state
maser emission from $^{28}$SiO toward six evolved stars.  Using the Very 
Large Array, we simultaneously observed the $v=0, J=1-0$, 43.4-GHz, 
ground-state and the $v=1, J=1-0$, 43.1-GHz, first excited-state transitions 
of $^{28}$SiO toward the oxygen-rich evolved stars IRC+10011, 
$o$~Ceti, W Hya, RX Boo, NML Cyg, and R Cas and the S-type star $\chi$ Cyg.  
We detected at least one $v=0$ SiO maser feature from six of the seven stars 
observed, with peak maser brightness temperatures ranging from 10000~K to
108800~K.  In fact, four of the seven $v=0$ spectra show multiple maser 
peaks, a phenomenon which has not been previously observed.  
Ground-state thermal emission was detected for one of the 
stars, RX Boo, with a peak brightness temperature of 200~K.  Comparing the $v=0$ 
and the $v=1$ transitions, we find that the ground-state masers are much weaker 
with spectral characteristics different from those of the first excited-state 
masers.  For four of the seven stars the velocity dispersion is smaller 
for the $v=0$ emission than for the $v=1$ emission, for one star the 
dispersions are roughly equivalent, and for two stars (one of which is RX Boo) 
the velocity spread of the $v=0$ emission is larger.  In most cases, the peak 
flux density in the $v=0$ emission spectrum does not coincide with the $v=1$ 
maser peak.   Although the angular resolution of these VLA observations were 
insufficient to completely 
resolve the spatial structure of the SiO emission, the SiO spot maps produced 
from the interferometric image cubes suggest that the $v=0$ masers are more 
extended than their $v=1$ counterparts.  

\end{abstract}


\section{INTRODUCTION}

Although the study of SiO maser emission toward long-period variable 
stars is an active field of research, very little work has been done 
with regards to the vibrational ground-state ($v=0$) SiO emission 
toward these evolved objects.  The question of the existence of 
observable ground-state SiO masers was first posed by \cite{BSLJ:75}.  
The ground-state lines of $^{28}$SiO and the much rarer $^{29}$SiO 
and $^{30}$SiO isotopic species have been observed 
in a number of single-dish surveys conducted primarily at 43 GHz 
($v=0, J=1-0$) and 86 GHz ($v=0, J=2-1$).  In the two isotopic species, 
this ground-state emission has been shown to be relatively weak 
unsaturated maser emission \citep{AB:92}.  For $^{28}$SiO the nature of the 
$v=0$ emission is not as clear.  Although \cite{BSLJ:75} may have detected a  
weak $v=0, J=2-1$ maser line toward the star VY CMa, the general consensus on 
the 86.8 GHz line is that it is thermal circumstellar SiO emission with 
broad parabolic or flat-topped line profiles \citep[see for example][]{BPGMD:86}. 
The $v=0,J=1-0$, 43.4-GHz line was similarly thought to be thermal in nature, however in 
surveys by \cite{JWWS:85} and \cite{JSWWG:91} several of the $v=0$
line profiles showed narrow emission spikes ($<1$~km~s$^{-1}$) superposed, 
in some cases, on broader pedestals of emission.  \citeauthor{JSWWG:91} 
suggested that these narrow spikes could be maser emission with linewidths 
comparable to or narrower than the thermal Doppler width.  
The data of \cite{JSWWG:91} also suggested a correlation 
between the intensities of the $v=0$ and $v=1$ lines providing 
additional evidence for ground-state maser emission and the possibility 
that the gas producing the two types of emission is spatially coincident.

To date, very little interferometric imaging of the $v=0$ emission toward
SiO maser emitting regions has been conducted.  The more abundant $^{28}$SiO 
species was observed toward the Orion IRc2 star-forming region in the 
$J=1-0$ transition \citep{CD:95} and the $J=2-1$ transition \citep{WPML:95}.
In each case, the detected emission was found to have both maser and thermal 
components.  Observations of the rarer $^{29}$SiO $v=0, J=2-1$ transition 
toward Orion \citep{BHL:98} also showed this blend of thermal and maser emission.
A comparison of the $^{28}$SiO $v=1, J=2-1$ and $^{29}$SiO $v=0, J=2-1$
maser spots maps lead \cite{BHL:98} to conclude that the two isotopic
species are not excited in the same gas layers.  

Toward late-type stars,
interferometric observations of the $v=0, J=2-1$ $^{28}$SiO 
emission have been conducted by \cite{LUCAS:92} and by \cite{SB:93}. 
No evidence of ground-state masers was found in either observation with the possible 
exception being an intense spike of emission toward the star R Cas \cite{LUCAS:92}.  
Prior to the work presented here, no interferometric studies of the $v=0, J=1-0$
$^{28}$SiO transition had been conducted.  We therefore sought to detect and
image the ground-state SiO maser emission toward 7 stars from the list
of possible detections in \cite{JSWWG:91}. 
 
\section{OBSERVATIONS AND REDUCTION \label{OBS}}

Observations of the $v=0, J=1-0$, 43.4-GHz and $v=1, J=1-0$, 43.1-GHz
transitions of $^{28}$SiO were performed using the Very Large
Array (VLA) in C-configuration.  The VLA is maintained and operated by the
National Radio Astronomy Observatory (NRAO)\footnote
{The National Radio Astronomy Observatory is a facility of the National 
Science Foundation operated under cooperative agreement by Associated 
Universities, Inc.}.  We observed the 7 stars and associated
extragalactic phase calibrators over three epochs.  The stars $o$~Ceti
and IRC+10011 (WX~Psc) were observed on August 4, 2001; the stars 
NML~Cyg, R~Cas, and $\chi$~Cyg were observed on August 8, 2001; and
the stars RX~Boo and W~Hya were observed on August 10, 2001.
Data were simultaneously in two bandpasses of bandwidth 6.25 MHz
centered on line rest frequencies
of 43122.808 MHz and 43423.858 MHz for the $v=1, J=1-0$ and $v=0, J=1-0$
transitions respectively.  The VLA correlator in normal mode produced 64
channels per IF with a channel spacing of 97.656 kHz ($\sim$0.68 km~s$^{-1}$).
 
Data were reduced using the standard routines within the Astronomical 
Image Processing System (AIPS).  For the first epoch, the absolute flux 
density scale was established using 3C84 assuming a flux density for this
source of $10.5$~Jy.  For the second and third epochs, both 3C84 and 3C286 were 
used for the amplitude calibration.  The 43-GHz flux density of 3C286 
was assumed to be $1.49$~Jy.  For each star observed, a nearby extragalactic calibrator 
was also observed in order to estimate the instrumental and atmospheric phase 
fluctuations. Phase corrections estimated from the calibration sources were 
applied to the target source data.  To eliminate ringing in channels adjacent 
to strong maser features, the data were Hanning smoothed prior to imaging.  
This resulted in a degradation of the spectral resolution by a factor of 2
to ($\sim$1.4 km~s$^{-1}$). 

For each source, an iterative self-calibration and imaging procedure 
was performed to map a single strong reference spectral feature in the $v=1$ 
transition.  The resulting phase solutions were applied to all channels 
in the the bands containing both the $v=1$ and $v=0$ transitions.
From the calibrated data, total intensity image cubes, consisting
of $256\times 256$ pixel images of the center 60 spectral channels, were 
produced for both the $v=1$ and $v=0$ SiO emission lines for each star.  
Naturally-weighted individual channel images were made with an approximately $23''\times 23''$ 
field of view.  The synthesized beam sizes ranged from $0.56''\times 0.54''$ for the
highest declination source, R~Cas, to $1.09''\times 0.43''$ for the lowest
declination source, W~Hya.  For the star RX~Boo, the $v=0$ emission 
was resolved out on long baselines.  We therefore reduced the resolution
for RX~Boo by a factor of two by limiting the $uv$ range used in the imaging.  This 
resulted in an image cube with a synthesized beam of $1.12''\times 1.09''$

Positions for 5 of the 7 stars as determined from our VLA observations are 
reported in Table \ref{REF_POS}.
Each position was determined from a two-dimensional (2-D) Gaussian fit to 
the peak in the image of the reference feature for the $v=1$ transition prior to 
self-calibration.  The errors in the position are based on the size of the 
synthesized beam divided by twice the signal-to-noise ratio (SNR) in the image.  
These errors represent a small fraction of the beam and are probably not 
representative of the true error in the derived position of the star.  There may
be significant blending of multiple maser features in the peak channel (discussed below).  
In addition, SiO masers typically exhibit ring-like structures
with shell sizes are on the order of tens of milliarcseconds.  The position of 
one or more maser features at random locations on the ring is therefore not
a good representation of the position of the stellar photocenter and may be off
by more than half the diameter of the SiO maser shell.  For the 
stars RX Boo and W Hya, the phase stability prior to self-calibration was 
insufficient to determine a reliable position from the image.  Therefore, 
no position is reported for these stars.     

Numerous Very Long Baseline Interferometry (VLBI) 
studies of circumstellar SiO masers have shown that a single spectral channel 
often contains multiple maser features; however, the angular resolution 
of our VLA data was not sufficient to resolve multiple features separated by 
distances less than typical maser shell diameters of 30--50~mas.  In order to 
parameterize individual maser components, 2-D
Gaussian functions were fit to the peak in each spectral 
channel of the image cubes using the AIPS task SAD.  The analysis of the 
spectral image cubes, the results of the Gaussian fits to the 
$v=0$ and $v=1$ SiO emission and the astrophysical interpretation are 
discussed below. 

\section{RESULTS}

\subsection{Ground-State Maser Emission}

\begin{figure}[hbt]
\plotone{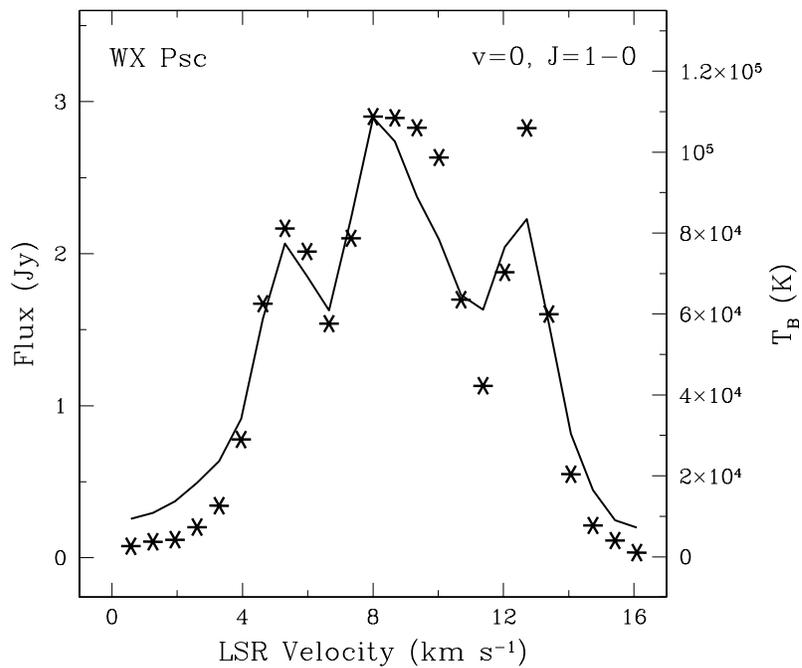}
\caption{Spectrum of the $v=0, J=1-0$ SiO emission toward IRC+10011 
(WX Psc) resulting from the 2-D Gaussian fits to the emission peaks 
in the VLA channel maps.  The solid line represents the flux density 
in each channel, while the points represent the corresponding brightness 
temperatures.  \label{WXPSC_TEMPB}}
\end{figure}

\begin{figure}[bth]
\plotone{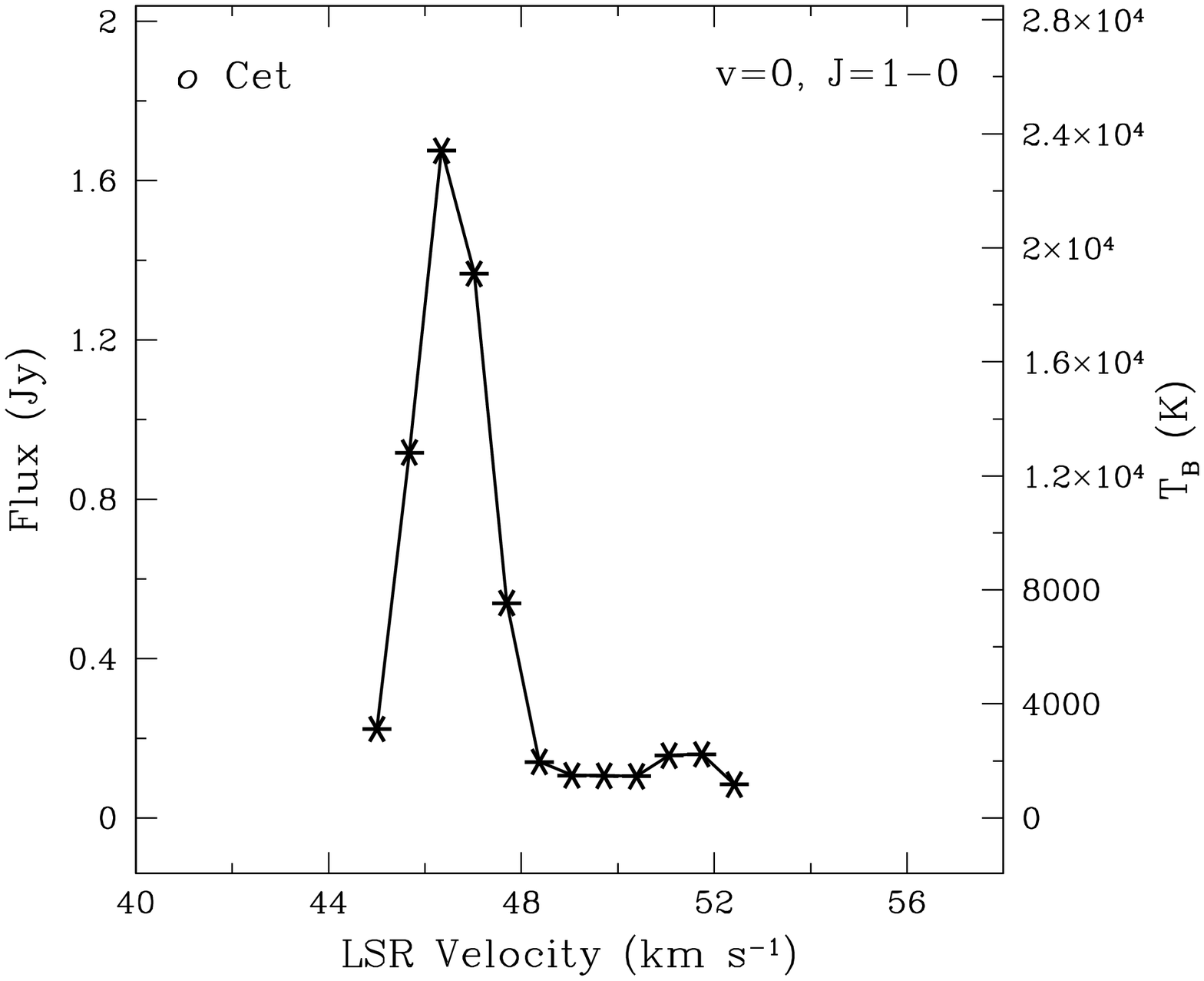}
\caption{Same as Figure \ref{WXPSC_TEMPB} for star $o$ Ceti.\label{MIRA_TEMPB}}
\plotone{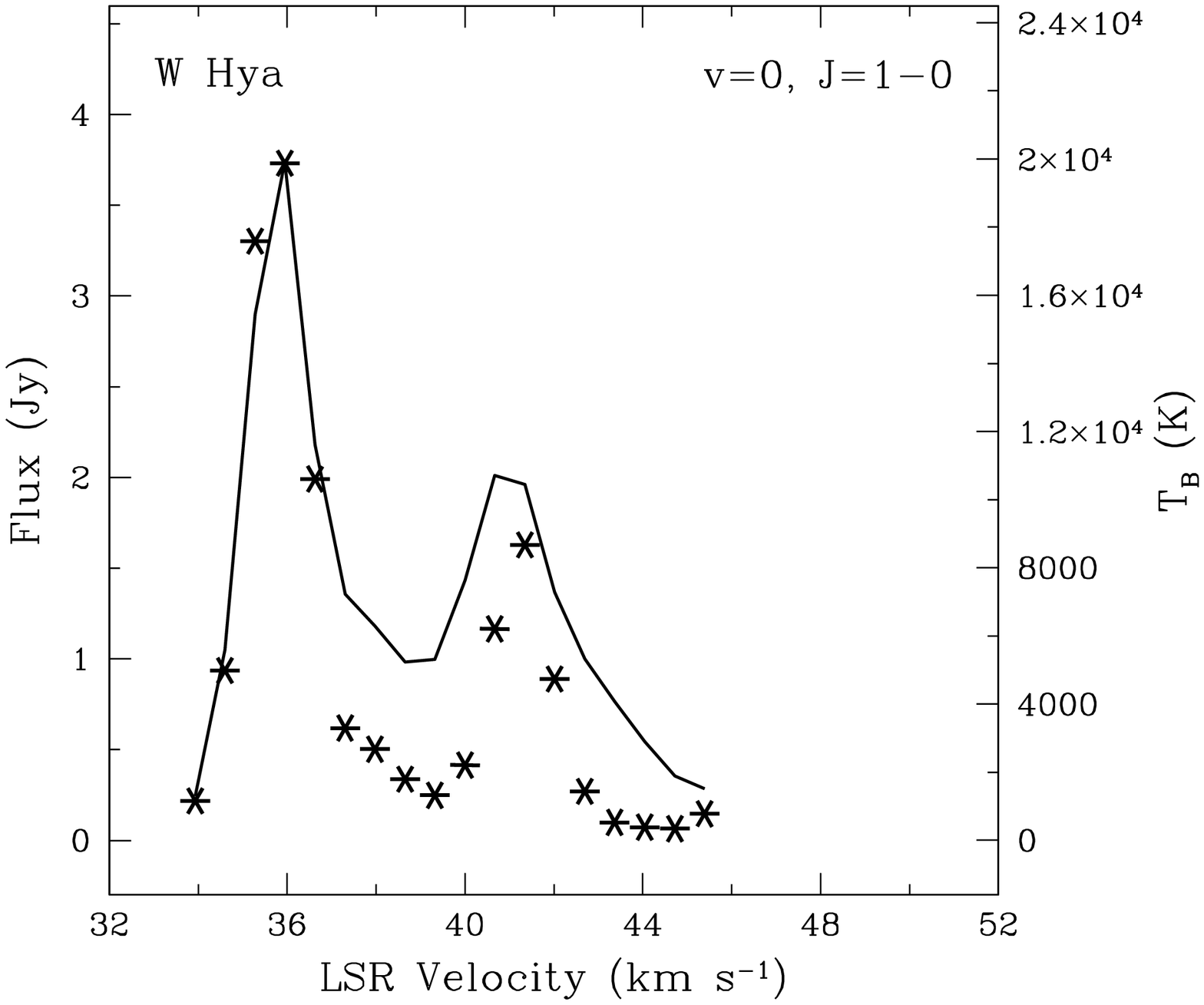}
\caption{Same as Figure \ref{WXPSC_TEMPB} for star W Hya.\label{WHYA_TEMPB}}
\end{figure}

\begin{figure}[hbt]
\plotone{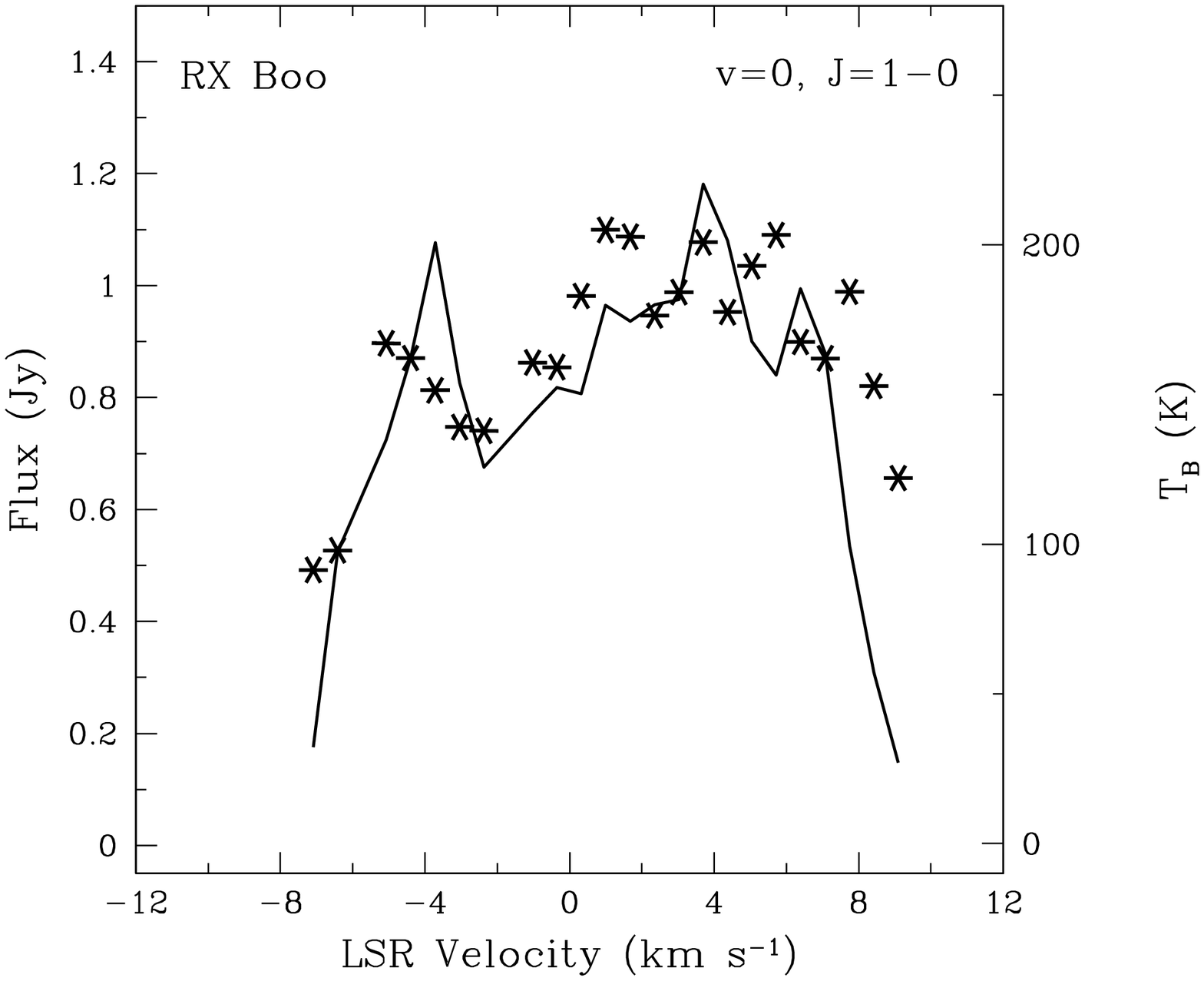}
\caption{Same as Figure \ref{WXPSC_TEMPB} for star RX Boo.\label{RXBOO_TEMPB}}
\plotone{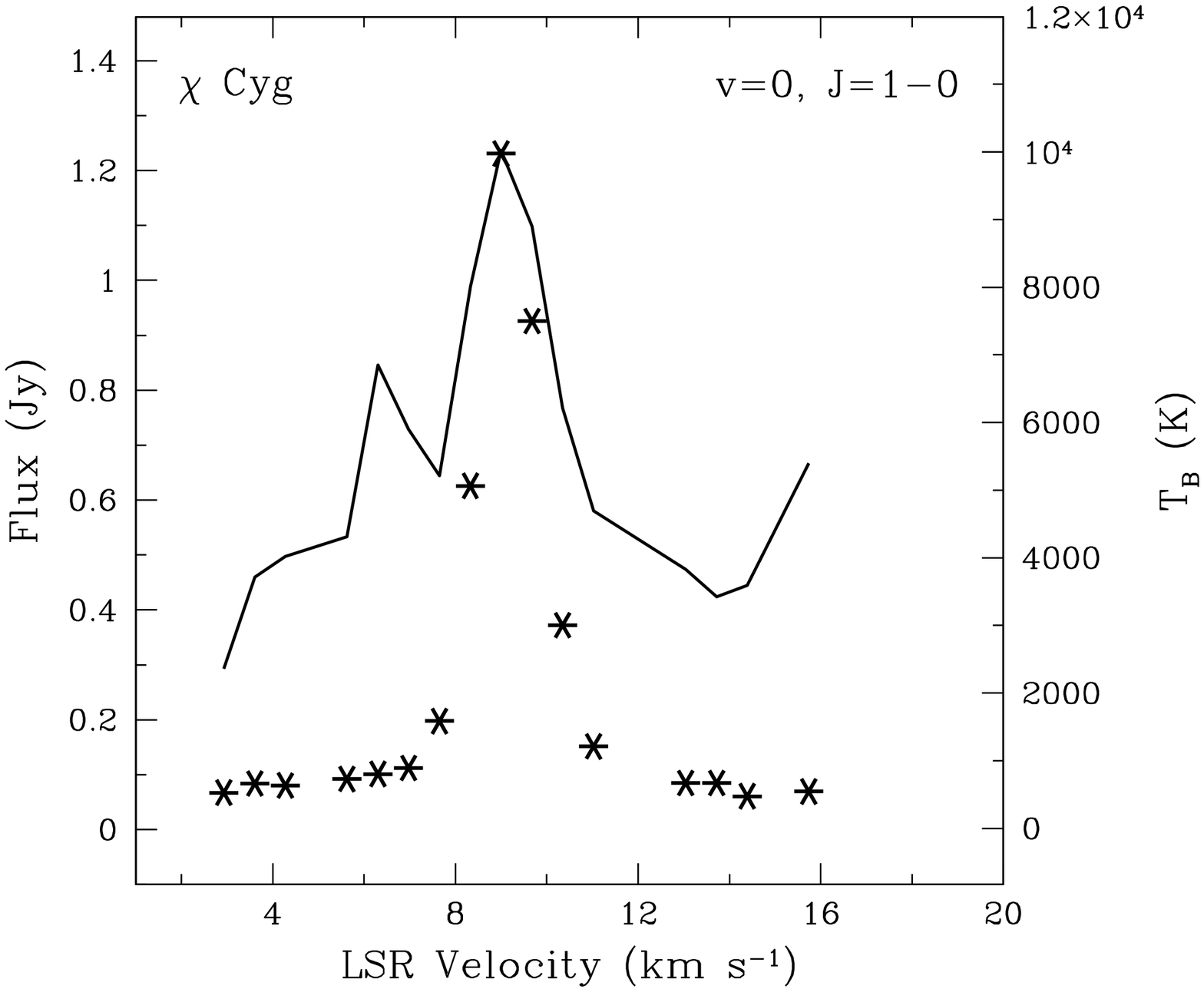}
\caption{Same as Figure \ref{WXPSC_TEMPB} for star $\chi$ Cyg.\label{CHICYG_TEMPB}}
\end{figure}

\begin{figure}[hbt]
\plotone{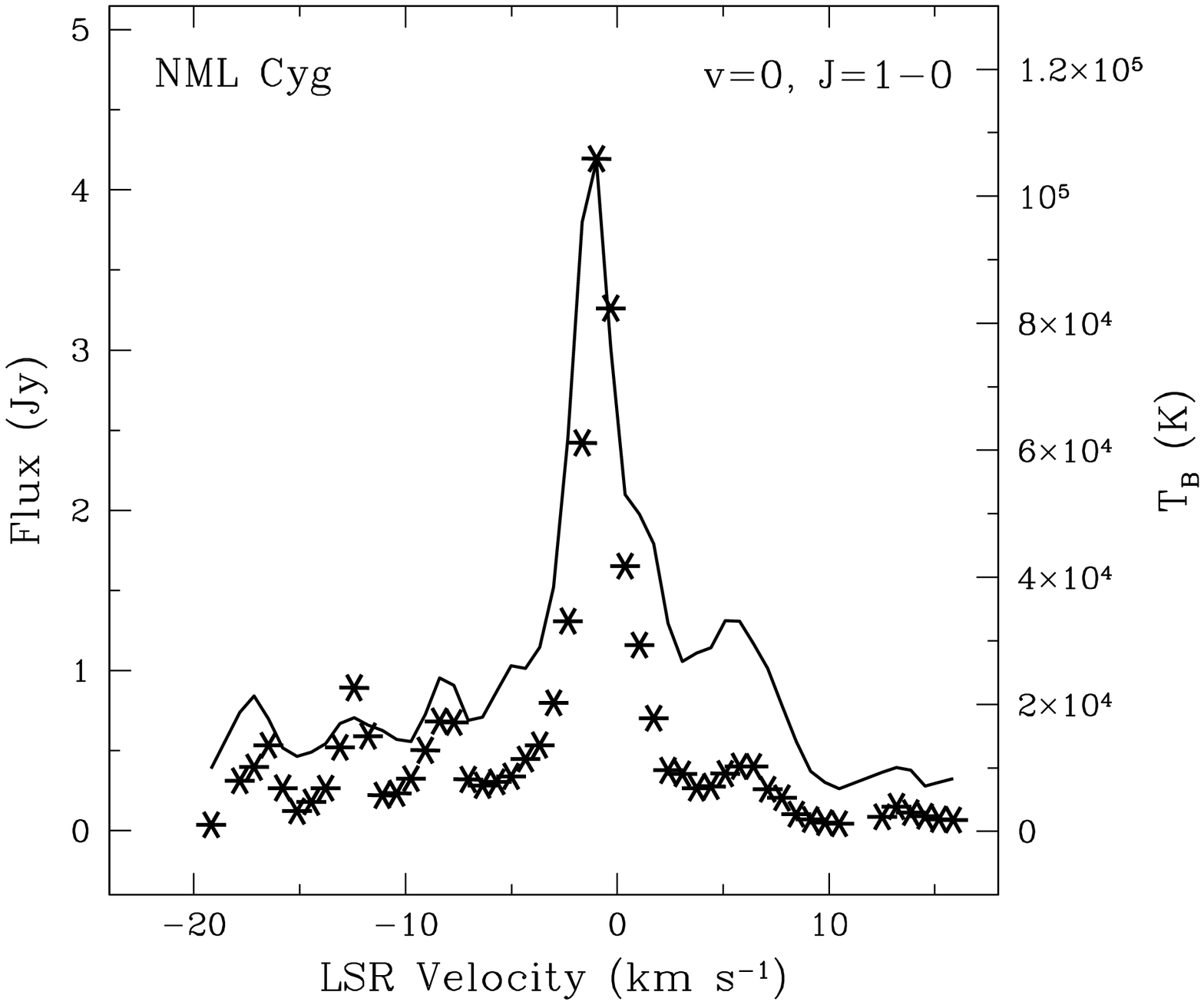}
\caption{Same as Figure \ref{WXPSC_TEMPB} for star NML Cyg.\label{NMLCYG_TEMPB}}
\plotone{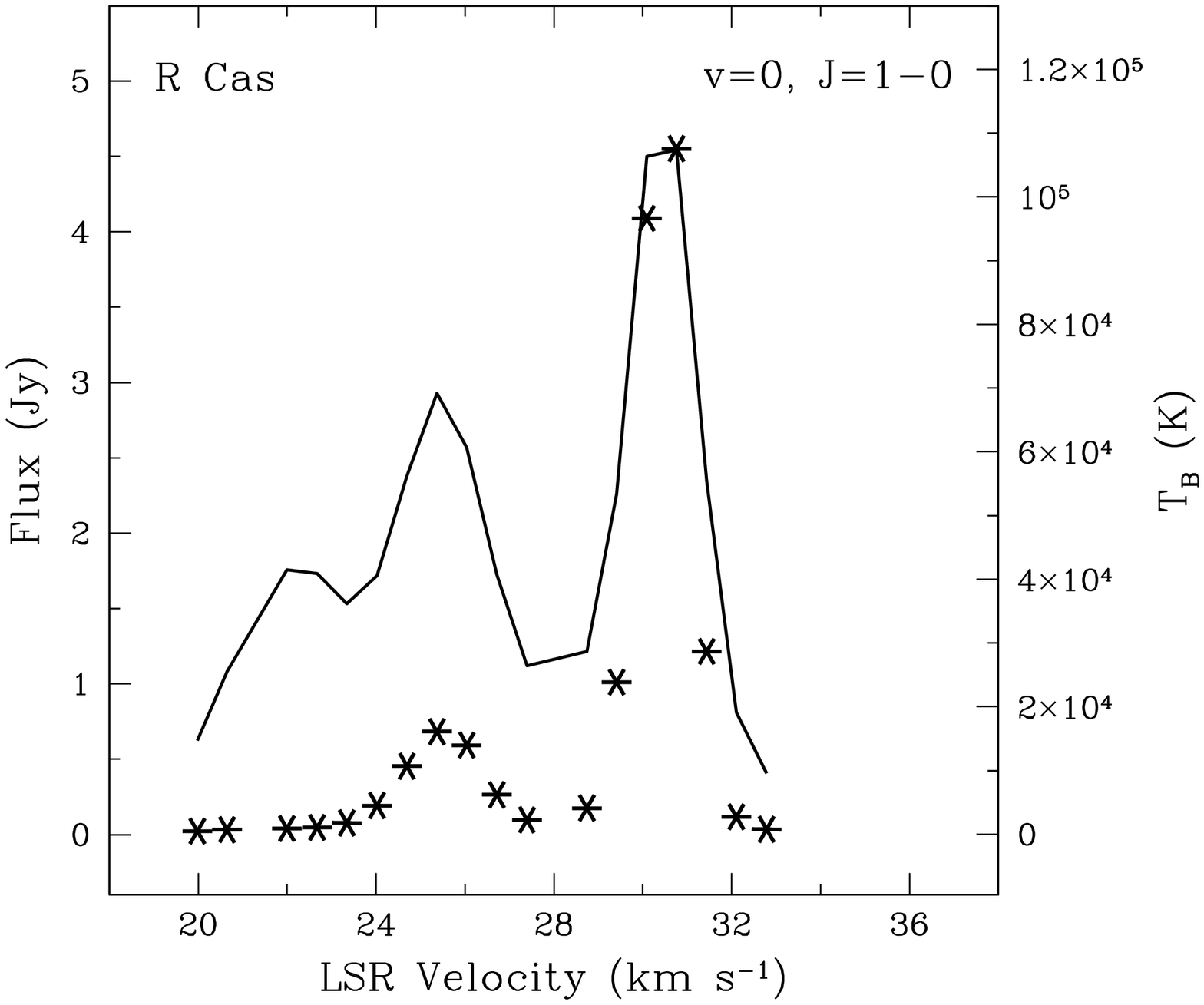}
\caption{Same as Figure \ref{WXPSC_TEMPB} for star R Cas.\label{RCAS_TEMPB}}
\end{figure}

As mentioned in the previous section, we fit 2-D Gaussians to the emission
peak in each spectral channel yielding component flux density, spatial position, 
and velocity information.  Since the primary goal of these observations was to 
determine whether ground-state SiO maser emission is exhibited in the 
atmospheres of late-type stars, we concentrate here on the ground-state,
$v=0$, transition.  In order to determine whether the observed $v=0$ SiO emission 
has a maser component to it, we converted the fitted component flux densities 
to equivalent brightness temperatures using:
\begin{equation}
T_b = \frac{S_{\nu} \lambda^2}{2 k \Omega_s} \approx 936 \frac{S_{\nu}}{{\theta_s}^2}
\end{equation} 
where, for the right-hand estimate,  $S_{\nu}$ is the flux density in Jy and 
$\theta_s$ is the geometric mean of the 
fitted source size in arcseconds.  Figures \ref{WXPSC_TEMPB}--\ref{RCAS_TEMPB} plot the 
flux densities and equivalent brightness temperatures for all 7 stars observed.  
In these figures, the solid line represents the fitted component flux densities in 
Janskys (Jy), and the points represent the equivalent component brightness 
temperatures.  For unresolved maser spots, we assumed source 
sizes of 0.5 times the beam.  The points are scaled by the logarithm of the flux 
density, and are color coded according to velocity bin.  The flux density, deconvolved 
source size, and brightness temperature for the strongest spectral feature toward 
each source are tabulated in Table~\ref{FLUX_TEMPB}.  

The kinetic temperature of the gas in the circumstellar envelope, $T_k$,can be 
estimated from
\begin{equation}
T_k = T_{\star} (\frac{R}{R_{\star}})^{-0.5}
\end{equation} 
where $T_{\star}$ is the stellar photospheric temperature, $R_{\star}$
is the stellar radius and R is the radial distance in the circumstellar
envelope (CSE) \citep{LW:84, HB:00}.
Assuming an estimate of $T_{\star} \approx 2500$~K, the kinetic temperature
in the region of the $v=1$ SiO masers is $\sim$1500~K.  We used this kinetic temperature
to compare to the brightness temperatures of the SiO ground-state emission, 
as a nominal test to determine whether the emission observed in the $v=0, J=1-0$ 
transition is indeed masing.  This temperature is rather conservative given the 
expectation that the $v=0$ masers should occur at a larger radial distance 
than the $v=1$ SiO masers (discussed later).  From Table~\ref{FLUX_TEMPB} 
we see that for 6 of the 7 stars observed, the peak brightness temperature is 
much higher than the nominal 1500~K kinetic temperature.  We take this comparison
to clearly indicate that some of the $v=0, J=1-0$ SiO emission is indeed maser
emission.  The only star that was observed that does not show maser emission from 
ground-state SiO in is RX~Boo.  In addition, the deconvolved angular sizes for the 
peaks in the six stars with suspected maser emission (see Table~\ref{FLUX_TEMPB}) 
were well below the corresponding beam size.  This is also suggestive of compact 
maser emission.  The only source with a deconvolved source size greater than the 
synthesized beam was again RX~Boo.

Examination of Figures \ref{WXPSC_TEMPB}--\ref{RCAS_TEMPB} shows that 4 of the 
6 ground-state maser sources show multiple brightness temperature peaks well above 
the nominal kinetic temperature.  The two exceptions are $o$~Ceti and $\chi$~Cyg. 
It is highly probably that these additional features are also ground-state masers.
Such multiple masing features contradict previously 
reported detections which showed single narrow emission peaks in the spectra.  
The six $v=0$ maser stars appear to show a range in the degree of masing
from $\chi$~Cyg and $o$~Ceti which exhibit single maser peaks to sources 
like NML~Cyg and IRC+10011 in which nearly the entire spectrum is above 
the brightness temperature threshold for masing.  The spectra for the latter 
two stars are similar to the $v=0, J=1-0$ spectrum for Orion IRc2, which \cite{CD:95} found
to be dominated by SiO maser emission.  For $\chi$ Cyg, it appears that the 
single maser feature sits on a broader pedestal of thermal SiO emission.  This is 
reminiscent of some of the $v=0, J=1-0$ spectra observed by \cite{JSWWG:91} 
and the $v=0, J=2-1$ spectrum for R Cas observed by \cite{LUCAS:92}.

\subsection{Comparison of the $v=0$ and $v=1$ SiO Emission}

\begin{figure}[hbt]
\plotone{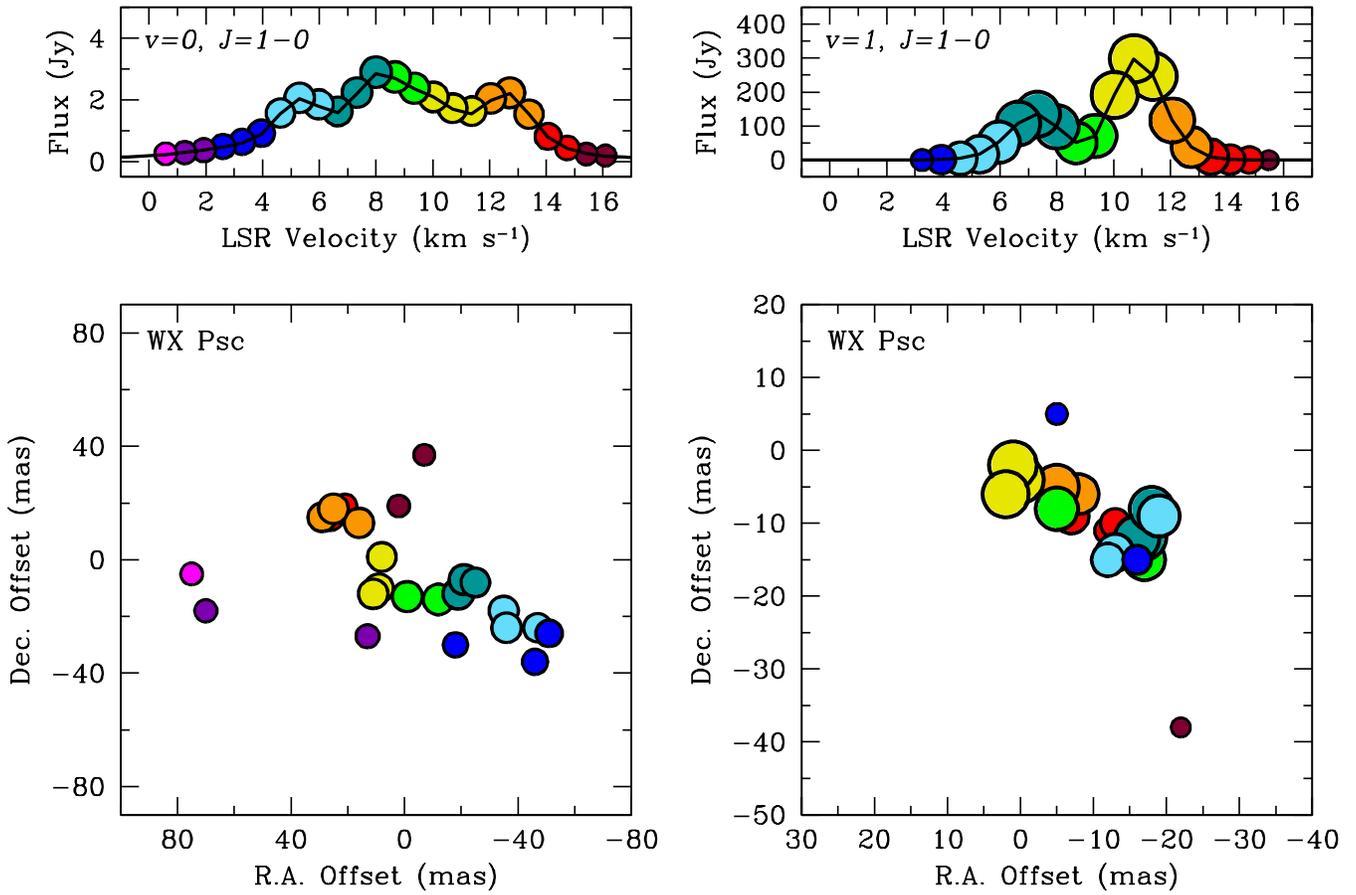}
\vspace{-1.5cm}
\caption{Spatial/velocity structure of the $v=0, J=1-0$ (left panels) and
$v=1, J=1-0$ (right panels) SiO emission toward IRC+10011 (WX Psc)
as measured by the 2-D Gaussian fits to the peaks in the VLA channel images.  
The two upper panels show the spectra formed by plotting flux density versus 
LSR velocity, color coded in velocity increments from blue-shifted (left) to 
red-shifted (right).  The two lower panels are spatial spot maps of the 
of the SiO emission.  The color of each point is representative of the corresponding 
velocity bin in the spectrum and the size of each point is proportional to the 
logarithm of the flux density.  The solid lines in the upper panels represent the 
vector-averaged, cross-correlated flux density in each channel averaged over
all antennas. \label{WXPSC_VEL}}
\end{figure}

\begin{figure}[hbt]
\plotone{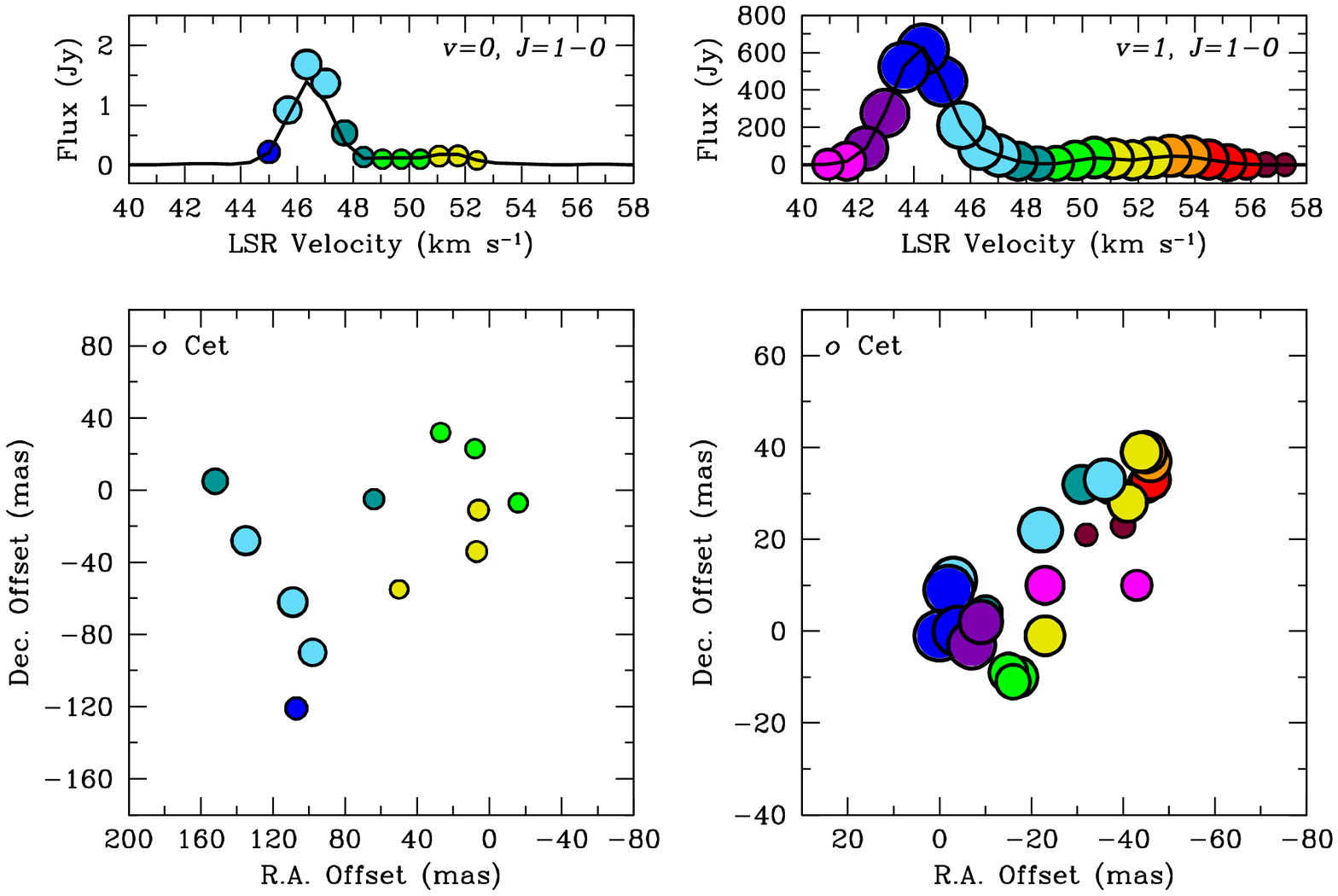}
\vspace{-1.5cm}
\caption{Same as Figure \ref{WXPSC_VEL} for star $o$ Ceti.\label{MIRA_VEL}}
\end{figure}

\begin{figure}[hbt]
\plotone{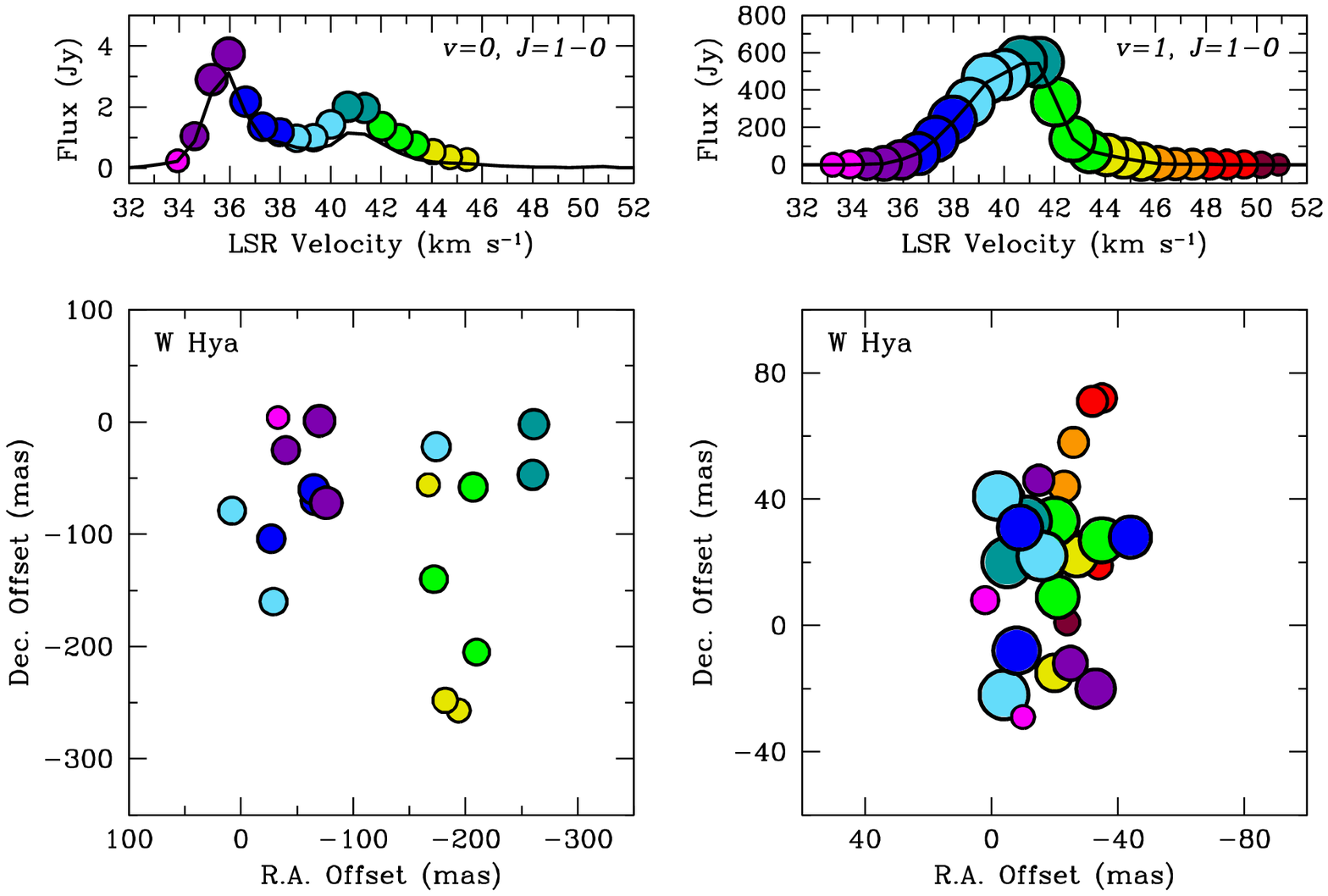}
\vspace{-1.5cm}
\caption{Same as Figure \ref{WXPSC_VEL} for star W Hya.\label{WHYA_VEL}}
\end{figure}

\begin{figure}[hbt]
\plotone{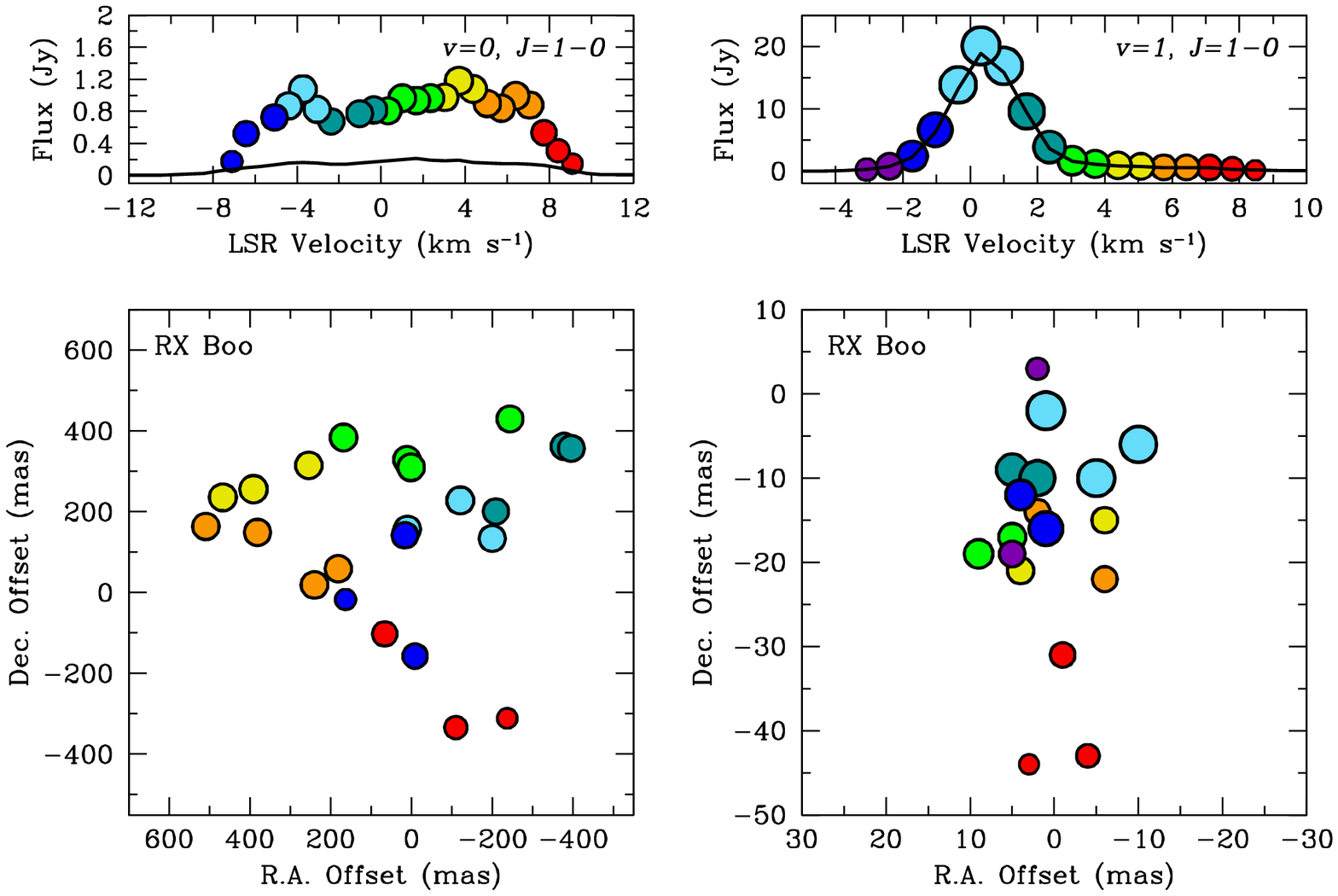}
\vspace{-1.5cm}
\caption{Same as Figure \ref{WXPSC_VEL} for star RX Boo.\label{RXBOO_VEL}}
\end{figure}

\begin{figure}[hbt]
\plotone{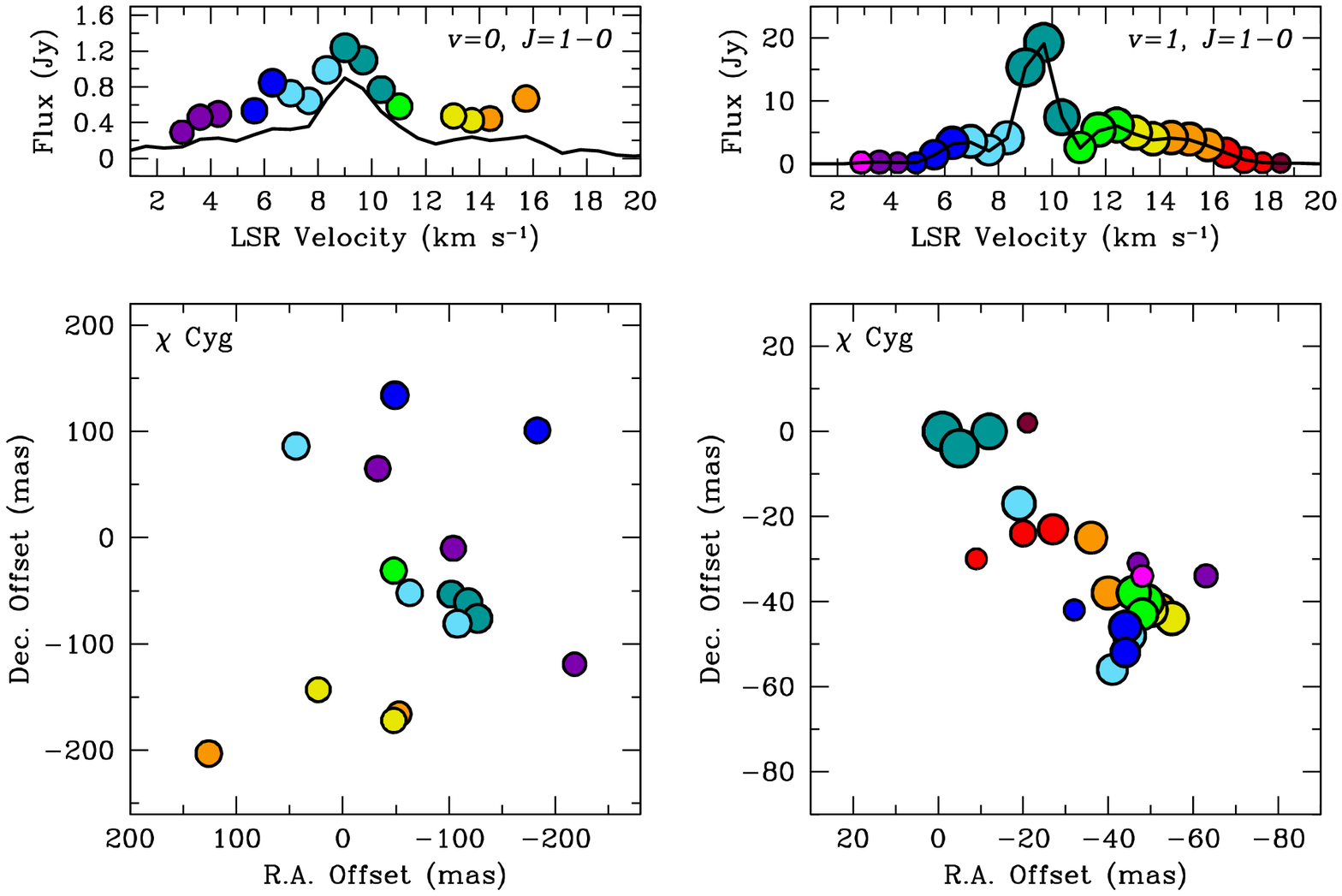}
\vspace{-1.5cm}
\caption{Same as Figure \ref{WXPSC_VEL} for star $\chi$ Cyg.\label{CHICYG_VEL}}
\end{figure}

\begin{figure}[hbt]
\plotone{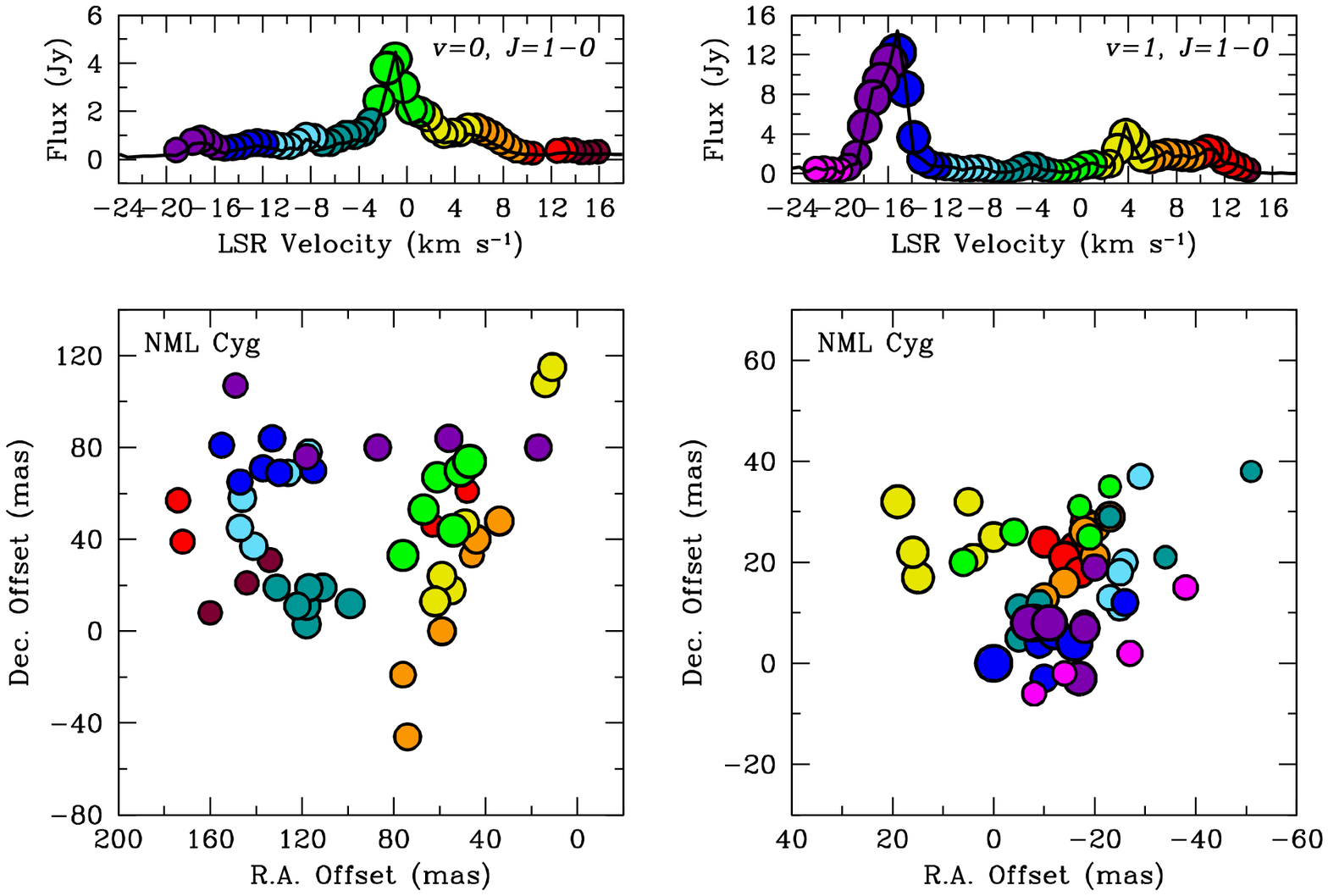}
\vspace{-1.5cm}
\caption{Same as Figure \ref{WXPSC_VEL} for star NML Cyg.\label{NMLCYG_VEL}}
\end{figure}

\begin{figure}[hbt]
\plotone{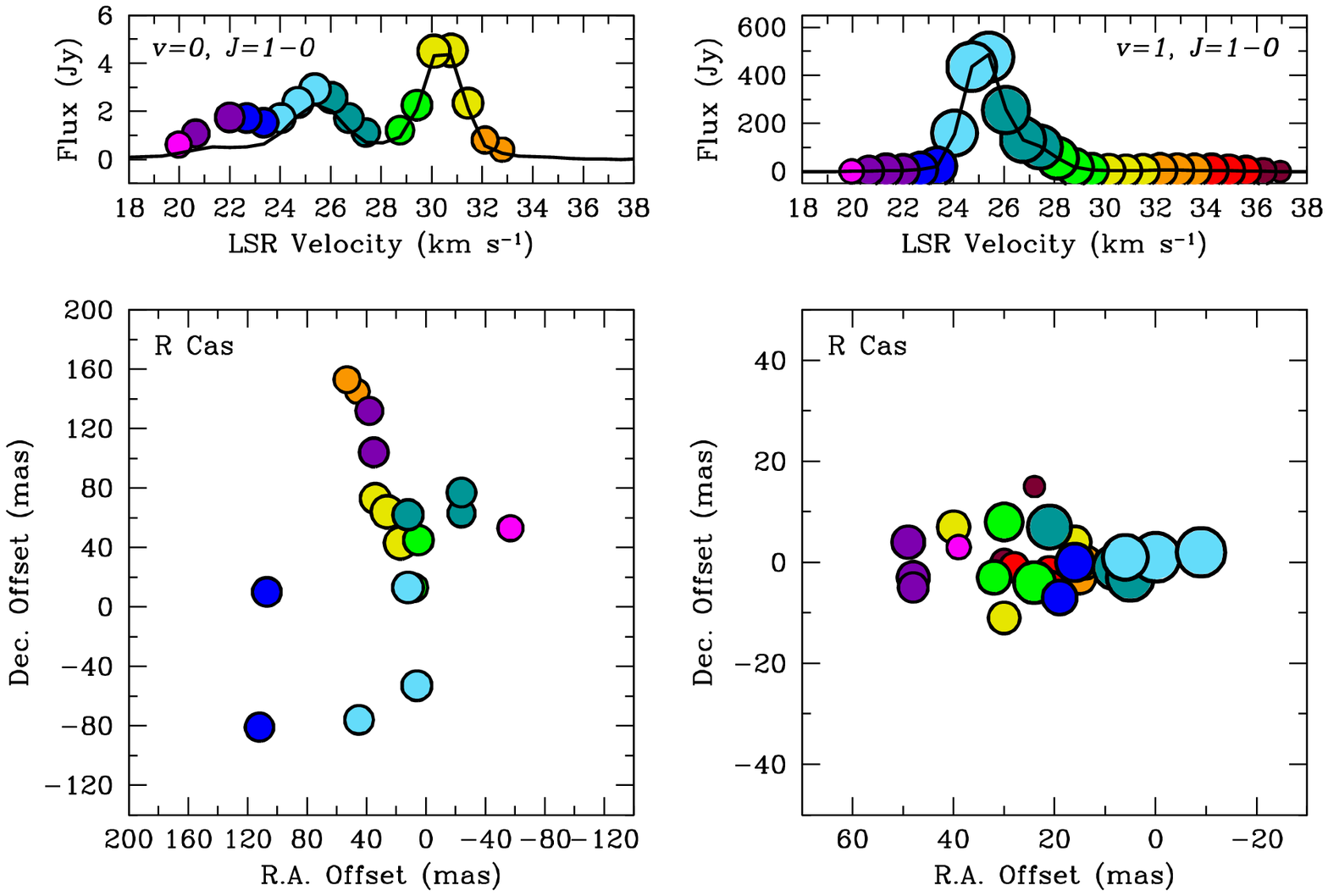}
\vspace{-1.5cm}
\caption{Same as Figure \ref{WXPSC_VEL} for star R Cas.\label{RCAS_VEL}}
\end{figure}

Having established that there is at least some degree of masing in the 
ground-state SiO emission toward 6 of the 7 stars observed, we sought to 
compare the $v=0$ emission with that observed for the more common 
$v=1, J=1-0$, first excited-state maser transition.  To facilitate this
comparison, we plotted the results of the 2-D Gaussian fits in 
Figures~\ref{WXPSC_VEL}--\ref{RCAS_VEL}.  In each figure, the two left panels 
show the spectrum (upper panel) and spatial distribution (lower panel) of 
the $v=0$ SiO emission.  Similarly, the two right panels show the spectrum 
and spatial distribution of the $v=1$ SiO emission.  Components derived 
from the 2-D fits are represented by color-coded circles with sizes 
scaled by the logarithm of the flux density and colors corresponding to
particular velocity bins.  The total velocity range and velocity 
binning is the same for each transition for each star.  The flux density
scale in the upper panels and the right ascension and declination offsets 
in the lower panels are adjusted to provide the best plot filling 
factor in each transition.  Also plotted in the upper panels of each 
figure, are solid lines which represent the vector-averaged, 
cross-power, spectrum averaged over all VLA antennas for comparison. 

Comparing the upper panels in Figures~\ref{WXPSC_VEL}--\ref{RCAS_VEL} we
notice that the spectra for the two SiO transitions toward each star are 
quite different.  Aside from the obvious flux density differences, the spectra
vary in number of peaks, center velocities of the peaks, and total velocity
dispersion in the transition.  In Table~\ref{SPEC_PROP} we list some of 
the relevant spectral parameters derived from the fits to the image data. 
Both the figures and the table show that for majority of the stars, 
$\chi$ Cyg being the one exception, the peak flux densities in the two 
transitions do not coincide in velocity to within 1~km~s$^{-1}$.  
Additionally, the velocity dispersions ($\Delta V$ in Table \ref{SPEC_PROP}) 
between the two transitions vary for each star.  In 4 of the 7 stars 
$\Delta V$ is larger for the $v=1$ transition than for the $v=0$ line.  
Two of the stars, IRC+10011 and RX Boo, have larger dispersions in the 
$v=0$ transition, although RX Boo is not masing in this transition, 
and one star, NML Cyg, has velocity dispersions which are roughly equivalent. 

Plotted in the lower panels of Figures~\ref{WXPSC_VEL}--\ref{RCAS_VEL} are
the spatial distributions of the SiO emission in each transition for each
star.  As mentioned in Section \ref{OBS}, the VLA in C-configuration
is insufficient to resolve multiple maser components within a spectral channel,
thus we fit a single 2-D Gaussian to the emission peak within each channel.
For the $v=1$ maser emission, the position errors reported by the AIPS task SAD 
for these fits range from $\sim$15~mas to less than 1~mas.  For the weaker 
$v=0$ maser emission, position errors ranged from $\sim$70~mas to 1 mas.  For 
the thermal $v=0$ emission toward RX Boo, errors ranged from 100 mas to 30 mas. 
These reported errors are based on the size of the beam divided by 
twice the signal-to-noise ratio (SNR) in the channel image and are considered 
more realistic than the formal errors of the Gaussian fit.  Nevertheless, the
reported errors are still a very small fraction (0.1--0.002) of the synthesized 
beam and some may question their reliability.  Because of this and the fact
that there is probably some component blending occurring in the images,  
the spatial distributions mapped in Figures~\ref{WXPSC_VEL}--\ref{RCAS_VEL} are 
discussed only in a qualitative sense.  A true quantitative comparison of the 
spatial distributions of the two masing transitions will have to wait for higher-resolution 
interferometric observations.   

With this in mind, we note that for all of the stars observed,  the spatial 
extent of the $v=0$ SiO emission is several times larger than that of the 
corresponding $v=1$ maser emission.  In addition, 
we find that the $v=1$ masers form compact distributions ranging from 
$\sim$30--90~mas consistent with typical SiO maser shell sizes as measured 
by VLBI.  Especially enticing are the maps for NML Cyg (see Figure \ref{NMLCYG_VEL}) in which 
we find that the $v=1$ maser distribution agrees with previous VLBI results 
from \cite{BM:00}.  Additionally, the $v=0$ ground-state masers appear to be resolved 
into a definite ring-like structure with a radius of about 50 mas.  Using the 
Berkeley Infrared Spatial Interferometer (ISI), \cite{DANCHI:01} determined the radius
of the inner dust shell of NML Cyg to be 80--100 mas.  This would place the 
$v=0$ masers in a region roughly half the radial distance between the $v=1, J=1-0$ masers 
and the inner circumstellar dust shell for NML Cyg.  Assuming a stellar size
of approximately 9~mas \citep{BM:00}, the kinetic temperature of the SiO in the
$v=0$ region is $\sim$750~K based on Equation 2.

\section{DISCUSSION}

Ground-state, $^{28}$SiO maser emission is rarely addressed in 
the literature in which numerical models for the generation of SiO masers are presented.  
This may be due to the fact that to date there has been very little evidence for 
the existence of masers in this transition.  
In an early paper, \cite{KS:74} commented that population inversions in the 
$v=0$ states are possible under certain conditions.  They stated that the 
percentage of the inversion is small due to the fact that the collision 
rate among the $v=0$ levels is comparable to the rate of radiative depopulation.
In one model consisting of strictly radiative excitation,\cite{KS:74} found 
an inversion of the $v=0, J=1-0$ line which occurs at a distance of 
$5\times 10^{14} - 4\times 10^{15}$~cm ($\sim$9--69 $R_{\star}$) with an 
emission rate down by a factor of 100 from the $v=1, J=1-0$ line.  

More recent investigations addressing the excitation of the SiO masers 
through overlaps with optical \citep{RKTP:96} and infrared \citep{HB:00} lines, 
demonstrated possible mechanisms for the generation of $v=0$, $^{28}$SiO masers.  
In the case of optical pumping, \cite{RKTP:96} found the ground-state emission 
to be thermal for the $v=0, J=2-1$ and $v-0, J=3-2$ transitions of $^{28}$SiO, 
but that maser emission was generated for the $v=0, J=1-0$
state in their simulations. The infrared pumping mechanism of \cite{HB:00} was used 
to resolve some of the previously unexplained detections of certain SiO maser transitions 
and to predict some possible new SiO masers.  In addition, their numerical model also
demonstrated that $v=0, J=1-0$ $^{28}$SiO could be generated.  In both 
the optical and infrared pumping cases, the $v=0$ masers are generated in 
regions of the CSE where the SiO abundances are lower than for the typical excited-state 
masers.  Since the SiO abundance decreases with distance from the star as SiO 
condenses onto dust grains, then the $v=0$ masers are naturally formed at distances 
beyond that of the vibrationally excited masers in these models.

Our qualitative comparison of the spatial distributions supports the idea that 
the $v=0$ SiO masers occur at distances greater than $v=1$ maser shell.  In 
this case the $v=0$ maser distribution is several times the diameter of the 
corresponding $v=1$ masers.   This contradicts the speculation 
by \cite{JSWWG:91} that the correlation between the intensities of the 
$v=0$ and $v=1$ lines suggests that emission from the two transitions is
spatially coincident.  \cite{BHL:98}, in their interferometric observations 
of the $v=0, J=2-1$ $^{29}$SiO and the $v=1, J=2-1$ $^{28}$SiO masers toward 
Orion IRc2, determined that the ground-state and excited-state masers were 
not co-spatial.  They state that this is a consequence of all collisional/radiative 
pumping models which predict that various vibrational state masers should peak 
in different spatial regions of the CSE.

\section{SUMMARY}

Interferometric observations using the VLA have allowed us to definitively detect
ground-state $v=0, J=1-0$ $^{28}$SiO masers toward six long-period variable 
stars, IRS+10011 (WX Psc), $o$~Ceti, W Hya, NML Cyg, R Cas, and $\chi$ Cyg.
Thermal $v=0$ emission was detected for the only other the star in our sample, 
RX Boo.   Peak brightness temperatures for the ground-state masers ranged from 
10000~K to 108800~K at the time of our observations.  Of the six ground-state 
maser stars, four exhibited multiple maser peaks, a phenomenon not 
previously observed.  

Comparing the $v=0$ transition with the simultaneously observed $v=1, J=1-0$
maser emission, we find that the ground-state masers are weaker with spectra 
different than the corresponding first excited-state masers.  In general the 
peak in the $v=0$ spectrum for a given star occurs at a different velocity
from that of the $v=1$ maser peak.  Additionally, the dispersions in velocity
are not the same.

We were able to make low-resolution image cubes of both transitions for each 
star from our VLA C-configuration data.  From these images we created spot maps 
of the emission.  Although the resolution of our observations was insufficient 
to provide a quantitative comparison of the spatial morphologies of the two 
transitions, we can say that the ground-state SiO masers appear to be distributed 
over a larger region of the CSE than their $v=1$ counterparts.  
Future higher-resolution observations should enable us to clearly 
resolve the spatial distribution of ground-state masers toward these 
long-period variables.  The precise placement of these masers in the 
circumstellar envelope relative to the $v=1$ masers and to the circumstellar
dust shell has important implications for circumstellar chemistry and maser
pumping mechanisms.

\clearpage

\begin{deluxetable}{lrcr}
\tabletypesize{\footnotesize}
\tablewidth{0pt}
\tablecaption{Positions of $v=1, J=1-0$ SiO maser reference features. \label{REF_POS}}
\tablehead{  & \colhead{$V_{\rm peak}$} & \colhead{$\alpha$ (J2000)} & 
\colhead{$\delta$ (J2000)} \\
Source  & \colhead{(km s$^{-1}$)}  & \colhead{(h m s)} & \colhead{($^{\circ}$ $'$ $''$)}  }
\startdata

IRC+10011  \dotfill  &    10.7  &  01 06 25.9886 $\pm$0.00004 ($\pm$0.0006$''$) &    12 35 53.0373 $\pm$0.0006 \\
$o$ Ceti   \dotfill  &    44.3  &  02 19 20.7909 $\pm$0.00003 ($\pm$0.0004$''$) & $-$02 58 39.8577 $\pm$0.0005 \\
W Hya      \dotfill  &    40.7  &  \nodata &  \nodata \\
RX Boo     \dotfill  &     0.3  &  \nodata &  \nodata \\
$\chi$ Cyg \dotfill  &     9.7  &  19 50 33.9177 $\pm$0.00022 ($\pm$0.0028$''$) &  32 54 50.5987 $\pm$0.0021 \\
NML Cyg    \dotfill  & $-$15.2  &  20 46 25.5430 $\pm$0.00007 ($\pm$0.0008$''$) &  40 06 59.4177 $\pm$0.0010 \\
R Cas      \dotfill  &    25.4  &  23 58 24.8805 $\pm$0.00015 ($\pm$0.0014$''$) &  51 23 19.7645 $\pm$0.0014 \\
\enddata
\end{deluxetable}

\begin{deluxetable}{lrrcr}
\tabletypesize{\footnotesize}
\tablewidth{0pt}
\tablecaption{Peak flux densities and brightness temperatures of $v=0, J=1-0$ 
SiO emission. \label{FLUX_TEMPB}}
\tablehead{       & \colhead{Peak Flux} &                         & \colhead{Angular} & 
\colhead{Brightness}  \\
                  & \colhead{Density}   & \colhead{Velocity}      &  \colhead{Size}   & 
\colhead{Temp.}  \\     
         Source  & \colhead{(Jy)}      & \colhead{(km s$^{-1}$)}  &  \colhead{($''$)} &  
\colhead{(K)} }
\startdata
IRC+10011  \dotfill  &  2.90$\pm$ 0.02  &     8.0  &  0.16  &  108800 \\
$o$ Ceti   \dotfill  &  1.68$\pm$ 0.02  &    46.4  &  0.26  &   23400 \\   
W Hya      \dotfill  &  3.74$\pm$ 0.05  &    36.0  &  0.42  &   19900 \\
RX Boo     \dotfill  &  1.81$\pm$ 0.05  &     3.7  &  2.34  &     200 \\
$\chi$ Cyg \dotfill  &  1.24$\pm$ 0.02  &     9.0  &  0.34  &   10000 \\
NML Cyg    \dotfill  &  4.18$\pm$ 0.03  &  $-$1.0  &  0.19  &  105900 \\  
R Cas      \dotfill  &  4.54$\pm$ 0.04  &    30.8  &  0.20  &  107600 \\  
\enddata
\end{deluxetable}

\begin{deluxetable}{lrrrrrr}
\tabletypesize{\footnotesize}
\tablewidth{0pt}
\tablecaption{Observed spectral properties of the $v=0$ and $v=1$ SiO emission. \label{SPEC_PROP}}
\tablehead{          & \multicolumn{3}{c}{$v=0, J=1-0$}  &  \multicolumn{3}{c}{$v=1, J=1-0$} \\
                     & \colhead{}              &   \colhead{}             & \colhead{Peak Flux} & 
\colhead{}           &   \colhead{}            & \colhead{Peak Flux}  \\
                     & \colhead{$\Delta V$}    & \colhead{$V_{\rm peak}$} & \colhead{Density}   &  
\colhead{$\Delta V$} & \colhead{$V_{\rm peak}$} & \colhead{Density}  \\     
         Source   & \colhead{(km s$^{-1}$)} & \colhead{(km s$^{-1}$)}  & \colhead{(Jy)}      &
\colhead{(km s$^{-1}$)} & \colhead{(km s$^{-1}$)}  & \colhead{(Jy)}  }
\startdata
IRC+10011  \dotfill  &  15.5 &     8.0  &   2.90$\pm$ 0.02  &   
                        12.2 &    10.7  & 298.0 $\pm$ 0.6     \\
$o$ Ceti   \dotfill  &   7.4 &    46.4  &   1.68$\pm$ 0.02  &    
                        16.3 &    44.3  & 617.6 $\pm$ 0.9     \\   
W Hya      \dotfill  &  11.5 &    36.0  &   3.74$\pm$ 0.05  &   
                        17.7 &    40.7  & 551.1 $\pm$ 4.4     \\
RX Boo     \dotfill  &  16.2 &     3.7 &    1.18$\pm$ 0.05  &    
                        11.5 &     0.3  &  20.1 $\pm$ 0.2     \\
$\chi$ Cyg \dotfill  &  12.8 &     9.0  &   1.24$\pm$ 0.02  &   
                        15.6 &     9.7  &  19.25$\pm$ 0.05    \\
NML Cyg    \dotfill  &  35.1 &  $-$1.0  &   4.18$\pm$ 0.03  &    
                        36.0 & $-$15.2  &  12.25$\pm$ 0.04    \\  
R Cas      \dotfill  &  12.8 &    30.8  &   4.54$\pm$ 0.04  &   
                        17.0 &    25.4  & 475.9 $\pm$ 1.9     \\  
\enddata
\end{deluxetable}

\clearpage

\epsscale{0.6}
\epsscale{1.0}

\end{document}